\documentclass[%
 reprint,
 amsmath,amssymb,
 aps,
pra,
]{revtex4-2}

\usepackage{graphicx}% Include figure files
\usepackage[justification=raggedright]{caption}
\usepackage{dcolumn}% Align table columns on decimal point
\usepackage{bm}% bold math
\usepackage[bottom]{footmisc}
\usepackage{epsfig,latexsym,cancel,amssymb,amsmath,bm}
\usepackage{graphicx, caption}
\usepackage{feynmp}
\usepackage{subfigure}
\usepackage{epstopdf}
\usepackage{color}
\usepackage{mathtools}
\usepackage{bigints}
\usepackage[utf8]{inputenc}
\usepackage[english]{babel}
\usepackage[dvipsnames]{xcolor}
\usepackage{placeins}

\graphicspath{{Images/}}

\newcommand{\be}{\begin{eqnarray}}
\newcommand{\ee}{\end{eqnarray}}

\def\to{\rightarrow}

\begin{document}

\preprint{APS/123-QED}
	
	\title{Instability-induced patterning of a jelling jet}
	%In-fluo printing of soft filaments of varying composition and geometry
	\author{Aditi Chakrabarti}
	\thanks{A.C. and S.A. contributed equally }
	\affiliation{Paulson School of Engineering and Applied Sciences, Harvard University, Cambridge, MA 02138, USA}
	%\thanks{These two authors contributed equally }
	
	\author{Salem Al-Mosleh}
	\thanks{A.C. and S.A. contributed equally }
	\affiliation{Paulson School of Engineering and Applied Sciences, Harvard University, Cambridge, MA 02138, USA}

	\author{L. Mahadevan}
	\email[e-mail: ]{lmahadev@g.harvard.edu }
	\affiliation{Paulson School of Engineering and Applied Sciences, Department of Physics, Department of Organismic and Evolutionary Biology, Harvard University, Cambridge, MA 02138, USA}

    \date{\today}

	\begin{abstract}

	When a thin stream of aqueous sodium alginate is extruded into a reacting calcium chloride bath, it polymerizes into a soft elastic tube that spontaneously forms helical coils due to the ambient fluid drag.  We quantify the onset of this drag-induced instability and its nonlinear evolution using experiments, and explain the results using a combination of scaling, theory and simulations.  By co-extruding a second (internal) liquid within the aqueous sodium alginate jet and varying the rates of co-extrusion of the two liquids, as well as the diameter of the jet, we show that we can tune the local composition of the composite filament and the nature of the ensuing instabilities to create soft filaments of variable relative buoyancy, shape and mechanical properties. All together, by harnessing the fundamental varicose (jetting) and sinuous (buckling) instabilities associated with the extrusion of a jelling filament, we show that it is possible to print complex three-dimensional filamentous structures in the ambient fluid. %PRL version 
	\end{abstract}
	
	\maketitle
		Slender filaments of solids and viscous fluids undergo buckling, folding as well as coiling spanning length scales of several orders of magnitude from the nanometric to the macrometric. For example, an elastic rope fed uniformly toward a horizontal plane coils into a spool  \cite{mahadevan1996coiling,habibi2007coiling,jawedReis2014}; similar behavior is seen with a slender jet of viscous fluid   \cite{taylor1969instability,mahadevan1998fluid,mahadevan2000correction,ribe2017liquid}. In both cases, the presence of the plane boundary leads to compressive stresses, while geometric scale separation implies the dominance of bending deformations. Quantitative predictions for the coiling rates and radii then arise from the balance of internal elastic or viscous forces and  inertia and/or gravity. The patterns arising from these instabilities have also been harnessed to fabricate structures by coupling them to phase changes in solidifying polymers and glass \cite{kim2010nanopottery,brun2017molten}.  Here, inspired by the quest to pattern filaments in three dimensions without a sacrificial scaffold as exemplified by the 3Doodler pen \cite{dilworth2015hand}, we ask if there are instabilities and patterns that can be harnessed when a filament moves through an ambient fluid while also changing from a liquid to solid. This is seen in high speed liquid polymer extrusion in air \cite{Rutledge}, where solvent-loss driven solidification happens so fast that it is almost impossible to control the instability. But could one extrude a viscous filament in an aqueous bath containing a crosslinker that causes the stream to polymerize, buckle and coil due to the resisting viscous drag forces from the surrounding medium, even in the absence of compressive stresses due to a solid boundary? And then control the patterns by varying the extrusion rate and even the properties by co-extruding a second fluid within the jelling jet?  Here, we use a simple alginate-calcium system to study the onset of a coiling instability as well as its saturation using a combination of scaling analysis, stability theory and computation.  By introducing a second fluid at the inlet, we show that we can add periodic droplet inclusions of oil and gas to the soft filaments. This allows for a single-step fabrication protocol to make a continuous soft filament of varying composition, stiffness, curvature and buoyancy that can be used to write and draw patterns in 3 dimensions in an ambient fluid without a sacrificial substrate.
		
\section*{Experimental observations}
\begin{figure*}
\includegraphics[width=17.8cm]{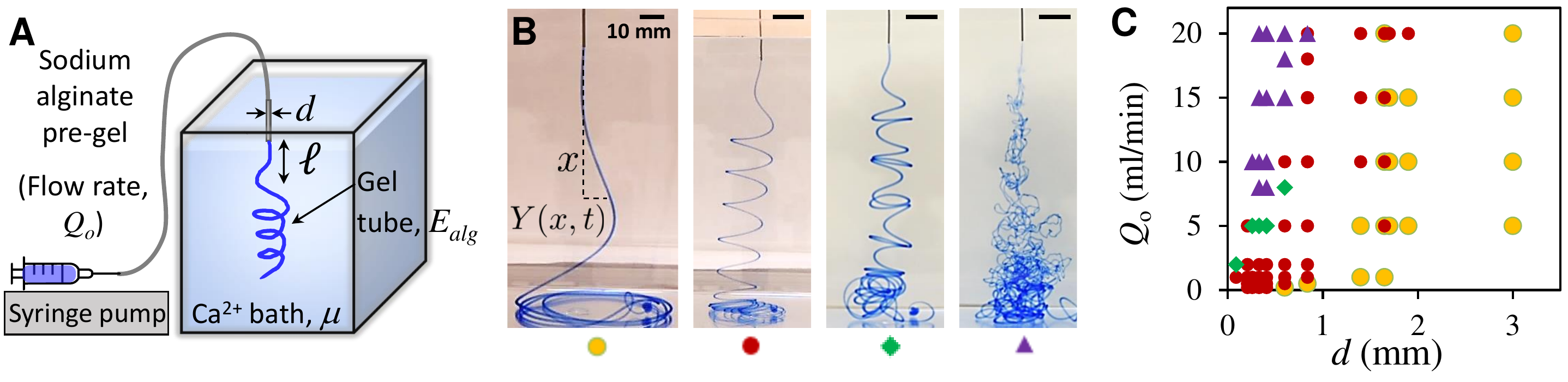}
\caption{\textbf{Instability of an extruded jelling jet.} (A) Schematic of the experimental setup that is used to obtain the drag induced coiling instability of a solidifying sodium alginate gel tube in a water bath containing 150 mM calcium chloride. A syringe pump is used to deliver the 2\% sodium alginate solution, dyed with methylene blue, at a prescribed flow rate ($Q_o$) via a blunt-tipped needle (inner diameter, $d$) into the calcium bath. As the sodium alginate solution contacts the calcium ions in the water bath, it begins to crosslink to form an elastic soft tube and then buckles due to the compression induced by the viscous drag of the surrounding liquid. (B) As the flow rate $Q_{o}$ is increased, we get a transition of behaviors from classical rope coiling (yellow circles, 18G needle, $d=0.84$mm at $Q_{o}$=0.5ml/min), to drag-induced elastic coiling (red circles, 22G needle, $d=0.41$mm at $Q_{o}$=0.5ml/min), period doubling (green diamonds, 22G needle, $d=0.41$mm at $Q_{o}$=5ml/min) to chaotic crumpling (purple triangles, 22G needle, $d=0.41$mm at $Q_{o}$=20ml/min). All scale bars are 10 mm. (C) Phase space of the instability regimes shown in (B) plotted as a function of the flow rates of the alginate extrudate $Q_{o}$ and the needle diameters $d$ used.}\label{figure_1}

\end{figure*}

Our experimental setup (Fig. \ref{figure_1}A) for the fabrication of single phase gel filaments and tubes uses a 2\% w/w sodium alginate solution, dyed with 0.01\% w/w methylene blue,  filled in plastic syringes, and dispensed inside an aqueous bath (viscosity $\mu$) of calcium chloride salt (150mM in water) at extrusion rates ($Q_{o}$, 0.1 - 20 ml/min) via needles (inner diameter $d$, 0.1-2mm) (Materials and Methods and Movie S1). For all the experiments, the concentrations of sodium alginate gel and the calcium ion concentration in the aqueous bath were kept fixed, thereby fixing the reaction rate ($k$) of gelation. The elastic Young's modulus of the crosslinked alginate tubes increases as the polymerization progresses, where it goes from hundreds of pascals (within a few seconds, \cite{ lee2012effect}) to tens of kilopascals (over several minutes), measured via gravity driven buckling of filament segments removed at different times from the calcium chloride solution (Materials and Methods, and SI Appendix, Fig.~S1~A-B).

For extrusion from a given needle diameter ($d$), at very low flow rates ($Q_{o}$)  a thin stream of alginate extruding inside the calcium bath polymerizes to form an elastic tube. In this regime, as the elastic tube touches the base of the container, it starts to coil like an elastic rope\cite{mahadevan1996coiling,habibi2007coiling}(Fig. \ref{figure_1}B, first panel), due to the compressive stresses induced in the filament when it contacts the floor. When $Q_{o}$ is increased, the polymerized elastic tube begins to buckle over a characteristic length $\ell$ much smaller than the falling height, and spontaneously forms loosely coiled helices (radius, $R$, spacing between coils, $\Lambda$) as it falls through the liquid bath (Fig. \ref{figure_1}B, second panel and SI Appendix, Fig.~S2). In this regime, there is no influence of the floor. Further increasing $Q_{o}$ leads to more complex patterns that exhibit signs of period doubling in the spacing between the helical coils(Fig. \ref{figure_1}B, third panel), and at very high extrusion rates, the jet polymerizes and crumples chaotically forming rough tubes (Fig. \ref{figure_1}B, last panel, SI Appendix, Movies S2-S3). The phase space of morphologies is determined by the extrusion rate ($Q_{o}$) and the needle diameter ($d$) as shown in Fig. \ref{figure_1}C (Movie S2 and a dimensionless phase space in SI Appendix, Fig.~S3). 

    \begin{figure*}[t!]
    \includegraphics[width=17.8cm]{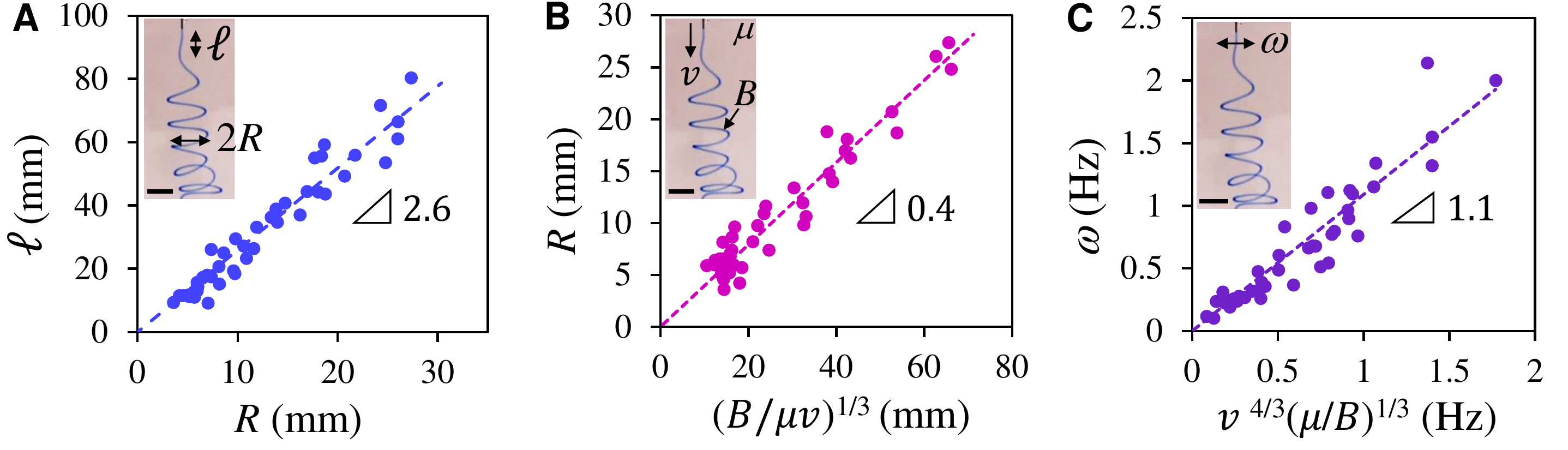}
    \caption{\textbf{Understanding the instability.} (A) The directly measured buckling length $\ell$ is plotted as a function of the coiling radius $R$ across various experiments, where needle diameters $d$ varied from 0.1 mm to 2 mm, and flow rates varied from 0.2 ml/min to 20 ml/min. $\ell$ shows an expected linear correlation with $R$ with a slope of 2.6. (Inset) Experimental snapshot to define $\ell$ and $R$. (B) Characterization of the drag induced elastic coiling instability via coiling radius ($R$) plotted as a function of the scaling relationship $(B/\mu v)^{1/3}$, with a slope of 0.4 for the linear regression. (Inset) Extrusion velocity $v$, viscosity of bath $\mu$ and bending stiffness $B$ of the polymerized filament. (C) The frequency of coiling $\omega$ for the experimental data shows that it scales as $v^{4/3}(\mu/B)^{1/3}$ with a slope of 1.1. (Inset) Definition of $\omega$. Scale bars in insets in (A-C) are 5 mm each.}\label{figure_2}
    \end{figure*}
\section*{Scaling analysis and stability theory} \label{sec:linTheory}

If the polymerization process with a rate $k$ is relatively fast, i.e. $kd/v \gg1$, where $v$ is the speed of extrusion and $d$ the diameter of the needle which is comparable to the diameter of the tube ($2r$) that forms, the polymerized filament behaves like an elastic filament of bending stiffness $B$. If the filament buckles over a characteristic scale $\ell$, we can balance the resisting bending torque $B/\ell$ in the filament with the forcing viscous torque from the ambient fluid that scales as $\mu v\ell^2$ to find a characteristic length scale \cite{machin1958wave,gosselin2014buckling}
\begin{equation}
              \ell \sim \left(\frac{B}{\mu \; v}\right)^{1/3},
\end{equation}
         above which the extrudate should buckle and coil into a helix spontaneously. Beyond the onset of buckling, we expect the radius of the helical coil to
follow this same scaling with $R \sim \ell$, while the frequency of coiling is expected to behave as \\
\begin{equation}
             \omega \sim v / \ell \sim v^{4/3} \;\left(\frac{\mu}{B}\right)^{1/3}.
        \end{equation}
        \indent %Substituting the typical experimental parameter values, $E \sim 2.7$ kPa, $r \sim 0.5$mm, $\mu \sim .001$Pa.s, $v \sim 15$mm/s, we get the buckling length $\ell \sim 22$mm and $\omega \sim 0.7$Hz. 
        Plotting the experimental data for buckling length $\ell$, coiling radius $R$ and coiling frequency $\omega$ (Fig.~\ref{figure_2}), we find that the three plots agree well with the scaling expressions obtained above. The vertical spacing between the coils ($\Lambda$) also remain steady while the stack of freshly coiled alginate tubes descends through the ambient calcium solution (SI Appendix, Fig.~S2 and Movie S1). Furthermore, we see that the entire range of scenarios from gravity-induced to drag-induced elastic coiling can be collapsed to a single universal scaling law that interpolates between the two limits and is given by ($\ell \sim (B/(\mu v + \Delta \rho g r^2))^{1/3}$) where $\Delta \rho$ is the density difference between the alginate tubes and the surrounding calcium bath (SI Appendix, Fig. ~S4). 
        %While the initial buckling of an elastic beam in a viscous medium has been looked into before \cite{gosselin2014buckling, kim2003macroscopic}, we investigated the non-linear regime of steady coiling in the medium, as well as different regimes of these instabilities due to viscous effects as a function of the filament extrusion rate.
        %
       
        Moving beyond a scaling analysis, we consider the linear stability of a thin elastic rod with length $L$, moving in a fluid with viscosity $\mu$ at a speed $v$, in the limit of low Reynolds number, i.e. $v L /\nu  \ll 1$.
        %so that the viscous drag force per unit to leading order in $r/L$ is given by \cite{cox_1970} $\mathbf{F}_{drag} = -\mu_s \left(2 \mathbf{V}^{\perp} + \mathbf{V}^{\parallel}\right),\;\;\;\;\;\; \mu_s \equiv \frac{2 \pi \mu}{\log(L/r)}$. Here $\mathbf{V}^{\parallel}$ and $\mathbf{V}^{\perp}$ are respectively the components of the velocity parallel and perpendicular to the rod. 
        For a flexible rod whose center line follows a path $Y(x,t)$ (see Fig. \ref{figure_1}B) that is only slightly deviated from being straight, force balance in the tangential and normal directions yields   
   \begin{eqnarray}
        0 &=& v \;\mu_{s} + \partial_x T(x, t), \nonumber \\
        0 &=& 2\; \mu_{s} \;\partial_t Y(x, t) + v \; \mu_{s} \; \partial_x Y(x, t) \nonumber \\ &+& B\; \partial_x^4 Y(x, t) - \partial_x  \left(T(x,t) \;\partial_x Y(x, t) \right). \label{eq:rod-in-flow}
   \end{eqnarray}
  Here $T(x,t)$ is the tension in the rod, and $\mu_s$ and $2 \mu_s$ are the effective drag coefficients in the tangential and normal directions ($\mu_s \sim \mu$, up to logarithmic factors in the aspect ratio) obtained from slender body theory \cite{cox_1970}. The accompanying boundary conditions are that at the free end $x = 0$, $\partial_x^2 Y(0, t) = \partial_x^3 Y(0, t) = T(0, t) =  0$ while at the needle $x = L = v t$, the clamped boundary conditions read $Y(L, t) = \partial_x Y(L, t) = 0$.
  
 Solving the first equation in Eq. [3] to get  $T = - \mu_s v x$ and substituting the result into the second Eq. in [3] while using the ansatz $Y(x, t) = \eta(x) e^{\sigma t}$ to determine the growth rate of an instability $\sigma$ yields the eigenvalue problem
   \begin{eqnarray}
        2  \mu_s \sigma \eta(x) +2 v \mu_s \eta'(x) + B \;\eta{''''}(x) + \mu_s  v  x \eta''(x) = 0.\;\;\;\;\;\; 
   \end{eqnarray}
   Since $\eta \to -\eta$ is a symmetry of the equation for real solutions $\eta(x)$, the growth rate $\sigma$ will also be real. At the onset of instability  ${\rm Re}\sigma =\sigma = 0$ so that Eq. [4] can be rewritten in scaled form as
   \begin{equation}
       2 \;\eta'(\xi) + \beta^{-1} \;\eta^{(4)}(\xi) +  \xi \;\eta''(\xi) = 0,\;\;\;\;\;\;
   \end{equation}
 where $x = L \xi$ and $\beta^{-1} \equiv B/(\mu_s v L^3)$. By using the boundary conditions at the free end, the solution to this equation can be written as 
 \begin{equation}
    \eta(\xi) =  A + C \; \xi \;\; _0F_1\left(\frac{4}{3}; -\frac{\xi^3 \; \beta}{9}\right), \label{eq:sol-rho}
 \end{equation}
 where  $_0F_1$ is the confluent hypergeometric function \cite{special-functions}.
 Using the boundary condition at the extruding end $\eta(1) = 0, \eta'(1)= 0$ then yields
 \begin{equation}
     _0F_1\left(\frac{4}{3}; -\frac{\beta}{9}\right) = \frac{\beta}{4}\;\;_0F_1\left(\frac{7}{3}; -\frac{\beta}{9}\right), \label{eq:boundary-cond1}
 \end{equation}
a condition only satisfied for special values of $\beta$, which then yields the smallest length at which the instability first occurs. Solving Eq.~\eqref{eq:boundary-cond1} numerically we find that this happens for $\beta \approx 3.48$ and thus yields   
 \begin{equation}
      \ell \approx 1.41 \left(\frac{B}{\mu v}\right)^{1/3} 
     %\left(\frac{3 \beta B}{2 \pi \mu v \;W(\frac{3 \beta B}{2 \pi \mu v r^3})}\right)^{1/3},
\end{equation}
%where $f(x)$ is a function that is solved numerically. In fact f(x) varies slowly and is nearly constant for our data set shown in Fig.~\ref{figure_1}. In particular, the mean is given by $\bar{f} \approx 1.41$ with coefficient of variation equal to $0.01$. Therefore, we expect $L_c \approx 1.41 \left({B}/{\mu v}\right)^{1/3}$. 
 This yields an expression for the oscillation frequency $\omega_c \equiv v/\ell \approx 0.71 \;v^{4/3} \left({\mu } / B\right)^{1/3} $. 
 
 Using experimental parameter values in the range $B \sim Er^4$ ($E=$2.7 kPa, $r \sim$ 0.1-2 mm), $\mu \sim$0.001 Pa.s and $v \sim$2-70 mm/s, the experimentally obtained relation for the buckling length is $\ell \sim 1.04 \left({B}/{\mu v}\right)^{1/3}$ 
 %(as $\ell \sim 2.6R$, and $R \sim 0.4 \left({B}/{\mu v}\right)^{1/3}$, 
 (Figs.~\ref{figure_2}A-B) while that for the frequency is $\omega \sim 1.1\; v^{4/3} \;\left({\mu}/{B}\right)^{1/3} $ (Fig.~\ref{figure_2}C), in reasonable agreement with the values obtained from the formal calculation above.
 % $W(y)$ is the solution to the equation $y = x \exp(x)$ for positive $x$. This function varies slowly and is nearly constant within our data set, $W^{1/3} \approx 2.35$. Therefore $L_c \approx 0.51 \left(\frac{B}{\mu v}\right)^{1/3}$. 
This linear stability calculation was confirmed using finite element simulations of elastic rods in a viscous bath (Materials and Methods, and Movie S4 and SI Appendix, Fig.~S5).

    \begin{figure*}[t!]
        \centering
        \includegraphics[width=17.8cm]{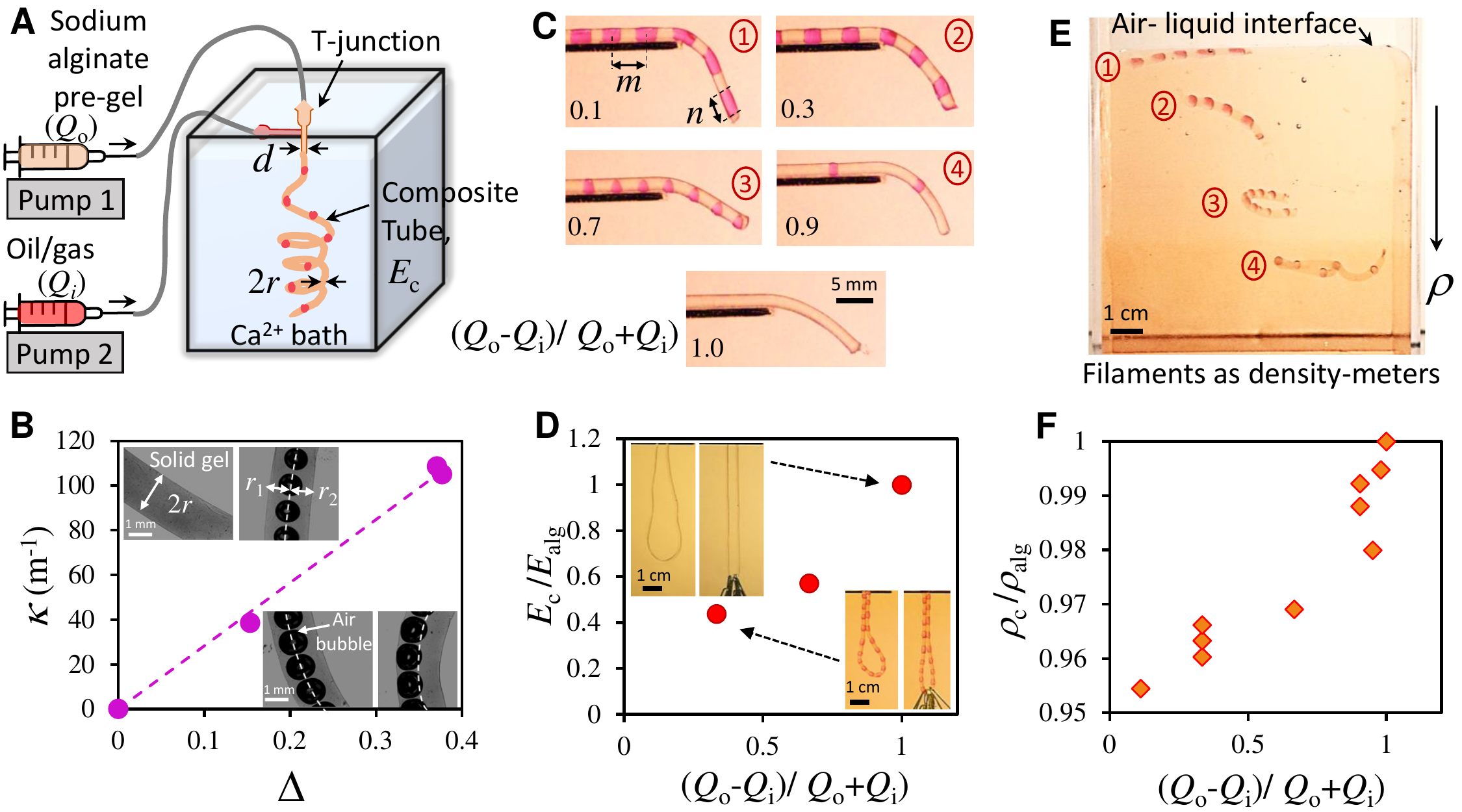}
        \caption{\textbf{Controlled patterning of co-extruded jelling jet.} (A) Schematic of the experimental setup that is used to obtain composite fluid-in-solid tubes via polymerization of co-extruding two liquids (2\% sodium alginate, $Q_{o}$ and inner fluid, $Q_i$) into an aqueous bath containing 150 mM calcium chloride (needle inner diameter, $d$). A T-junction in the tubing allows the alginate solution and embedded fluid to meet, before entering the calcium bath. (B) Curvature can be controlled by the inclusion of air bubbles at a prescribed (mis)alignment with the central axis of the tube, quantified by $\Delta = (r_1-r_2)/r$, which dictates composite tube curvature ($\kappa$). \textbf{  (C-D) Controlling elastic moduli with inclusions:} (C) Gravity driven bending of small segments of composite filaments show that elastic moduli of filaments vary when oil droplets (silicone oil, $\nu_s \sim $100 cSt, $\rho_{oil}=972$kg/m$^3$, flow rate $Q_i$, dyed red) are co-extruded within the alginate (outer extrusion rate, $Q_o$).  Varying the relative flow rates of oil and alginate $(Q_o-Q_i)/(Q_o+Q_i)$ controls filament morphology (spacing between oil inclusions, $m$ and the size of the inclusions, $n$) and therefore elastic moduli. (D) The normalized elastic moduli of the composite tubes ($E_c/E_{alg}$) can be obtained by stretching the tubes (inset) and plotting the normalized ratio $E_c/E_{alg}$ as a function of  $(Q_o-Q_i)/(Q_o+Q_i)$. \textbf{(E-F) Controlling density with inclusions: }(E) The density of alginate tubes (pure, $\rho_{alg}$=1054 kg/m$^3$) can be lowered by introducing oil inclusions ($\rho_{oil}=910$ kg/m$^3$) to form composite tubes ($\rho_c$, 1: 1006 kg/m$^3$; 2: 1018 kg/m$^3$; 3: 1021 kg/m$^3$; 4: 1045 kg/m$^3$). Suspending these filaments in a density gradient column serve as a direct readout for the density ($\rho_c$) of the column as the filaments become neutrally buoyant at a height where their density is equal to the density of the liquid at that height. Here, the density at the bottom of the container is 1050 kg/m$^3$ and at the air liquid interface is 1010 kg/m$^3$. (F) The normalized density of the composite tubes ( $\rho_c/\rho_{alg}$) are plotted as a function of  $(Q_o-Q_i)/(Q_o+Q_i)$. }\label{figure_3}
    \end{figure*}

 \section*{Controlling filament morphology and properties} \label{sec:curvature}

Using our understanding of the drag-induced coiling instability, we show how to harness it to control filament curvature. We first vary the flow rate $Q_{o} \sim v d^2$ and polymerization rate $k$; since the coiling is due to the ambient drag its curvature can get frozen in due to polymerization, depending on the scaled polymerization rate $kd/v$. At low flow rates, the tubes are relatively straight, while at higher $Q_{o}$, the tubes are strongly crumpled within an approximately conical plume, with an opening angle $\theta$ that varies linearly with needle diameter(SI Appendix, Figs.~S6, ~S7 and Movie S3). In this strongly crumpled regime, the polymerized tubes have a disordered morphology due to their hydrodynamic interaction with the bath even as they polymerize. This can be quantified via the variance of the curvature of the center line of these tubes as a function of the flow rate $Q_{o}$ (SI Appendix, Fig.~S7~E), which shows that as $Q_{o}$ is increased, the tubes go from being gently sinuous to highly rough. Turning this bug into a feature suggests a method to fabricate filaments of varying curvature by controlling the needle diameter and the extrusion rate, which 'freezes' the shape of the instability of the filament. In the chaotic regime, the small scale roughness in the filaments (SI Appendix, Fig.~S7) could also serve as a read out of the interfacial shear instability between the viscoelastic sodium alginate jet and the Newtonian fluid aqueous calcium chloride bath.
%
%    \section{Fluid-embedded soft tubes as deployable structures} \label{sec:fluidSolid}
%

To control specific properties of the elastic filaments such as buoyancy, rigidity, and curvature, we now introduce a second fluid, either oil (here, silicone, $\rho_{oil} \sim$ 910 kg/m$^3$), or gas (here, compressed nitrogen) that is internally co-extruded with the alginate solution, as shown schematically in Fig. \ref{figure_3}A (see also SI Appendix, Movie S5). At low co-extrusion rates, the internal fluid pinches off due to interfacial forces, and forms small droplet inclusions. By varying the relative flow rates of the co-extruding fluid ($Q_{i}$) and the encasing alginate ($Q_{o}$), we can control the size, spacing and position of the fluid droplet inclusions in the gel filaments. This allows us to control filament curvature by (mis)aligning the gas bubbles relative to the axis of the filament (Fig. \ref{figure_3}B). By varying the fraction of internal droplets of oil or gas, we can also control the effective filament bending and stretching stiffness (Figs. \ref{figure_3}C-D and SI Appendix, S1~C-D)) as well as the filament density and buoyancy (Figs. \ref{figure_3}E-F) (see also SI Appendix, Movie S6). { In addition, by varying the relative flow rates, we can alter the natural vibration frequency and internal damping of the filament (SI Appendix, Fig. S8)}. All together, this allows us then print and write out stable structures in the ambient liquid by dynamically varying the co-extrusion rate of the internal fluid (oil/gas) and the extrusion of the alginate, as seen in Fig.  \ref{figure_3}. 
        
 \section*{Conclusions}
      We have shown that it is possible to simultaneously harness the simplest instabilities in a slender elastic filament, that of buckling, and its analog in a slender liquid jet, that of break-up, to create a quasi-one dimensional liquid-solid composite. The co-extrusion a thin stream of aqueous sodium alginate and a second liquid into a reacting calcium chloride bath leads to the the pinch off of the internal liquid to form a beaded filament that buckles and coils. By controlling the relative rates of polymerization of the sodium alginate, with calcium ions concentration, and the mechanical instability of the freshly formed filaments, we have shown how one can use the underlying dynamical processes to to write in three dimensions while controlling filament properties such as buoyancy, stiffness, curvature as well as chirality. Our ability to control and steer the shape of the soft filaments using an instability stands in sharp contrast with the deterministic approach used to deposit each voxel in the additive manufacturing process \cite{skylar2019voxelated} and suggests alternative ways to manufacture and engineer low-dimensional fluid-solid composites \cite{style2020solid,bartlett2020introduction,cai2019unbounded} for various applications \cite{onoe2013metre,hu2010stretchable}.

\section*{Materials and Methods}{
\footnotesize
%Please describe your materials and methods here. This can be more than one paragraph, and may contain subsections and equations as required. 
\subsection*{Materials}
The pre-gel alginate solution was prepared by dissolving 2\% by weight (w/w) sodium alginate (Sigma Aldrich) in freshly collected deionized (DI) water, to which, 0.01\% weight (w/w) of methylene blue dye (Sigma Aldrich) was added to aid visualization. The aqueous calcium bath was prepared by dissolving 150mM of calcium chloride salt in DI water (i.e, 17g of CaCl$_2$ hexahydrate, Sigma Aldrich, in 1000g water), which were filled in clear plastic containers (10cm x 10cm x 12.5cm height). The dynamic viscosity $\mu$ of calcium bath was 0.001 Pa.s. Direct measurement of densities ($\rho$) gave 1054 kg/m$^3$ for 2\% sodium alginate and 1051 kg/m$^3$ for 150mM calcium chloride solution, and these concentrations of solutions were specifically chosen such that sodium alginate was marginally denser than the calcium bath. The 2\% sodium alginate solution was filled in plastic syringes, delivered via a syringe pump and dispensed inside the calcium bath via blunt-tip needles (32G to 12G, inner diameter $d$ ranging from 0.09mm to 1.9mm respectively). The extrusion flow rates $Q_o$ of the alginate solution was varied between 0.1 ml/min to 20 ml/min. We used new needles for each experiment as it was crucial to maintain a clean unclogged tip. Between each experiment, the calcium bath was stirred to maintain homogeneous ion concentration. The calcium bath was replaced by fresh solution after approximately ten experiments. For the multiphase composite tubes, embedding fluids such as silicone oil (density, $\rho_{oil}=910$kg/m$^3$, kinematic viscosity, $\nu$= 100 cSt) with red dye (silcPig), ferrofluid (FF310, Educational Innovations) and compressed nitrogen gas were injected at a flow rate of $Q_i$ (Fig. \ref{figure_3}A). By controlling relative rates of $Q_o$ and $Q_i$, various properties (buoyancy, curvature, stiffness) of the resulting composite fluid-in-solid filaments were controlled. All the experiments were performed at room temperature of 23$^o$ C. For thicker needles (12G-15G) a 25L plastic tank (41cm x 28cm x 30cm height) was used for the experimental measurements of the coiling instability.

\subsection*{Elastic moduli of the polymerized composite tubes}
The elastic modulus of the extruded polymerized tubes and filaments were measured by bending experiments of the filaments. Small segments ($\sim$ 1-2 cm) of the polymerized filaments were mounted such that its one edge was on a horizontal substrate. The other end was suspended such that gravity could weigh it down. The downward displacement ($\delta$) as a function of the overhung length of the filament from the edge ($\lambda$) were measured (SI Appendix, Fig. S1). A simple calculation using Euler-Bernoulli beam theory gives the solution for the bending profile of the elastic filament due to a distributed load $q$ (force per unit length). In particular, we get
\begin{eqnarray}
    \delta = \frac{q \; \lambda^4}{8 E I}, \;\;\;\; I = \frac{\pi\; r^4}{4}, \;\;\;\; q  \equiv \pi r^2 \;\rho \; g \implies \lambda^4 = \frac{2 E r^2}{\rho g} \delta.\;\;\;
\end{eqnarray}
Using the slope of the plot of $\lambda^4$ vs $\delta$ for freshly crosslinked alginate tubes extracted from solution close to the needle tip (seconds after it has formed, SI Appendix, Fig. S1~A) and away from the needle tip (several minutes after it has formed, SI Appendix, Fig. S1~B), the elastic modulus of the filament $E$ can be calculated from the slope of the linear regime $\frac{2 E r^2}{\rho g}$ (SI Appendix, Figs. S1~A-B). Similarly, this experimental setup was used to compare elastic moduli and effective stiffness of fluid-in-solid composite tubes (SI Appendix, Fig. S1~C). The elastic moduli of the composite tubes were estimated in stretching mode as well, where a filament segment was bent in a 'U' shape and vertically hung from a rigid boundary, and equal weights (metal pins) were placed in the neck of the 'U'. The elastic moduli were then estimated from the slope of the stress ($\sigma$)-strain ($\epsilon$) curves generated from the stretching experiments of the filaments (Figs.\ref{figure_3}C-D,  SI Appendix, Figs. S1~C-D and Movie S7). 
%The nominal stress $\sigma$ was calculated by the force applied divided by the original cross section of the alginate tubes before stretching. The strain $\epsilon$ was calculated as $\Delta L/L$, where $L$ was the original length before stretching and $\Delta L$ is the change in the length of the filament due to stretching.

\subsection*{Simulation}

	The discretized model of a filament involves a set of vertices connected by springs with rotational springs at each vertex (apart from the two boundary vertices) to account for resistance to bending (SI Appendix, Fig. S5~A). The discretized energy for the chain is given by
	\begin{eqnarray}
	   H = K_B \;N  - K_B\sum_{i= 1}^{N} \cos(\theta_i) + \frac{1}{2}K_S  \sum_{i = 0}^{N} \left(l_i - \bar{l}_i \right)^2,
	\end{eqnarray}
	where $l_{i} \equiv |\mathbf{e}_i| \equiv |\mathbf{r}_{i + 1} - \mathbf{r}_{i}|$ is the length of the edge and $\bar{l}_{i}$ is its rest length. For small angles the bending energy reduces to $K_B \theta_i^2/2$. We neglect twisting modes since for very thin rods, the speed of twist waves is much higher than for (long wavelength) bending waves and therefore reach equilibrium much faster. We can calculate the force on each vertex by taking the derivative $\mathbf{F}_{i} \equiv -d H/d \mathbf{r}_i$. We get
	\begin{eqnarray}
	    &\mathbf{F}_{i}& = -K_S \; \left[\left(l_{i-1} - \bar{l}_{i - 1}\right) \hat{\mathbf{t}}_{i - 1} -\left(l_i - \bar{l}_i\right) \hat{\mathbf{t}}_i \right] - \; \nonumber\\
	    &K_B& \left[\frac{P(\hat{\mathbf{t}}_{i - 2}, \hat{\mathbf{t}}_{i - 1})}{l_{i - 1}}  + \frac{P(\hat{\mathbf{t}}_{i}, \hat{\mathbf{t}}_{i - 1})}{l_{i - 1}} - \frac{P(\hat{\mathbf{t}}_{i - 1}, \hat{\mathbf{t}}_{i})}{l_{i}} -
	    \frac{P(\hat{\mathbf{t}}_{i + 1}, \hat{\mathbf{t}}_{i})}{l_{i}}\right],\;\;\;\;\;\;\;
	\end{eqnarray}
	where $\hat{\mathbf{t}}_{i} \equiv \mathbf{e}_i/l_i$ is the unit tangent along the edges and $P(\hat{\mathbf{t}}_{i}, \hat{\mathbf{t}}_{j}) \equiv \hat{\mathbf{t}}_{i} - (\hat{\mathbf{t}}_{i}\cdot \hat{\mathbf{t}}_{j}) \;\hat{\mathbf{t}}_{j}$ gives the component of $\hat{\mathbf{t}}_{i}$ perpendicular to $\hat{\mathbf{t}}_{j}$. \\
	The drag force for a slender rod can be estimated from Ref. \cite{cox_1970} and to leading order is given by
	\begin{equation}
	    \mathbf{F}^{drag}_{i} = -\mu_s \; \left(2 \mathbf{V}^{\perp}_i + \mathbf{V}^{\parallel}_i\right) \;\Delta{l}_{i}, %\;\;\;\;\;\;\;\;\;\;\;\;\;\; \mu_s = \frac{2 \pi \mu}{\log(L/r)}, 
	\end{equation}
	where $\mathbf{V}_i$ is the velocity of the vertex relative to the fluid and $\Delta{l}_{i}$ is the length of the Voronoi cell associated with the vertex. Implementing this simulation in C\texttt{++} code, we first perform a series of checks by comparing the results of the simulation to analytical calculations for beam buckling, gravity loading, and the known results for the spool-coiling of a rope falling under gravity onto the ground and get results consistent with theory and experiments \cite{mahadevan1996coiling,habibi2007coiling,jawedReis2014} (SI Appendix, Fig. S5~B and Movie S4). \\
	\indent In order to simulate the extrusion process, we add vertices at the extruding end (inlet point in SI Appendix, Figs. S5~B-D) at regular time intervals. The vertices are created attached elastically to the rest of the rod and moving with velocity $v$. By using experimentally motivated values for parameters ($E = 4.8$kPa, $Q = 10$ml/min, $d = 1$mm, $\Delta \rho = 15$kg/m$^3$.), we find that the onset of coiling behavior is analogous to the experimental observations with the coiling radius of the same order of magnitude as given in Fig. \ref{figure_2}B. However, the behavior in the simulation beyond the onset of the instability is different from experiments as shown in (SI Appendix, Fig. S5~D). In particular during experiments, coils are produced regularly with fixed radius as happens initially in the simulation. But over time the simulation coils start to unravel at the bottom free end. This discrepancy is likely due to the fact that our minimal model ignores nonlocal hydrodynamic interactions between elements of the rope and the gradual increase of rigidity along with a build up of natural curvature in the coils over time as further crosslinking occurs.
}
~\\\\
\footnotesize
\textbf{ ACKNOWLEDGMENTS.}{  For partial financial support, we thank the National Science Foundation grants NSF DMR 20-11754, NSF DMREF 19-22321, and NSF EFRI 18-30901. }

% Bibliography
\bibliography{references}

%apsrev4-2.bst 2019-01-14 (MD) hand-edited version of apsrev4-1.bst
%Control: key (0)
%Control: author (8) initials jnrlst
%Control: editor formatted (1) identically to author
%Control: production of article title (0) allowed
%Control: page (0) single
%Control: year (1) truncated
%Control: production of eprint (0) enabled
\begin{thebibliography}{22}%
\makeatletter
\providecommand \@ifxundefined [1]{%
 \@ifx{#1\undefined}
}%
\providecommand \@ifnum [1]{%
 \ifnum #1\expandafter \@firstoftwo
 \else \expandafter \@secondoftwo
 \fi
}%
\providecommand \@ifx [1]{%
 \ifx #1\expandafter \@firstoftwo
 \else \expandafter \@secondoftwo
 \fi
}%
\providecommand \natexlab [1]{#1}%
\providecommand \enquote  [1]{``#1''}%
\providecommand \bibnamefont  [1]{#1}%
\providecommand \bibfnamefont [1]{#1}%
\providecommand \citenamefont [1]{#1}%
\providecommand \href@noop [0]{\@secondoftwo}%
\providecommand \href [0]{\begingroup \@sanitize@url \@href}%
\providecommand \@href[1]{\@@startlink{#1}\@@href}%
\providecommand \@@href[1]{\endgroup#1\@@endlink}%
\providecommand \@sanitize@url [0]{\catcode `\\12\catcode `\$12\catcode
  `\&12\catcode `\#12\catcode `\^12\catcode `\_12\catcode `\%12\relax}%
\providecommand \@@startlink[1]{}%
\providecommand \@@endlink[0]{}%
\providecommand \url  [0]{\begingroup\@sanitize@url \@url }%
\providecommand \@url [1]{\endgroup\@href {#1}{\urlprefix }}%
\providecommand \urlprefix  [0]{URL }%
\providecommand \Eprint [0]{\href }%
\providecommand \doibase [0]{https://doi.org/}%
\providecommand \selectlanguage [0]{\@gobble}%
\providecommand \bibinfo  [0]{\@secondoftwo}%
\providecommand \bibfield  [0]{\@secondoftwo}%
\providecommand \translation [1]{[#1]}%
\providecommand \BibitemOpen [0]{}%
\providecommand \bibitemStop [0]{}%
\providecommand \bibitemNoStop [0]{.\EOS\space}%
\providecommand \EOS [0]{\spacefactor3000\relax}%
\providecommand \BibitemShut  [1]{\csname bibitem#1\endcsname}%
\let\auto@bib@innerbib\@empty
%</preamble>
\bibitem [{\citenamefont {Mahadevan}\ and\ \citenamefont
  {Keller}(1996)}]{mahadevan1996coiling}%
  \BibitemOpen
  \bibfield  {author} {\bibinfo {author} {\bibfnamefont {L.}~\bibnamefont
  {Mahadevan}}\ and\ \bibinfo {author} {\bibfnamefont {J.~B.}\ \bibnamefont
  {Keller}},\ }\bibfield  {title} {\bibinfo {title} {Coiling of flexible
  ropes},\ }\href@noop {} {\bibfield  {journal} {\bibinfo  {journal}
  {Proceedings of the Royal Society of London. Series A: Mathematical, Physical
  and Engineering Sciences}\ }\textbf {\bibinfo {volume} {452}},\ \bibinfo
  {pages} {1679} (\bibinfo {year} {1996})}\BibitemShut {NoStop}%
\bibitem [{\citenamefont {Habibi}\ \emph {et~al.}(2007)\citenamefont {Habibi},
  \citenamefont {Ribe},\ and\ \citenamefont {Bonn}}]{habibi2007coiling}%
  \BibitemOpen
  \bibfield  {author} {\bibinfo {author} {\bibfnamefont {M.}~\bibnamefont
  {Habibi}}, \bibinfo {author} {\bibfnamefont {N.}~\bibnamefont {Ribe}},\ and\
  \bibinfo {author} {\bibfnamefont {D.}~\bibnamefont {Bonn}},\ }\bibfield
  {title} {\bibinfo {title} {Coiling of elastic ropes},\ }\href@noop {}
  {\bibfield  {journal} {\bibinfo  {journal} {Physical Review Letters}\
  }\textbf {\bibinfo {volume} {99}},\ \bibinfo {pages} {154302} (\bibinfo
  {year} {2007})}\BibitemShut {NoStop}%
\bibitem [{\citenamefont {Jawed}\ \emph {et~al.}(2014)\citenamefont {Jawed},
  \citenamefont {Da}, \citenamefont {Joo}, \citenamefont {Grinspun},\ and\
  \citenamefont {Reis}}]{jawedReis2014}%
  \BibitemOpen
  \bibfield  {author} {\bibinfo {author} {\bibfnamefont {M.~K.}\ \bibnamefont
  {Jawed}}, \bibinfo {author} {\bibfnamefont {F.}~\bibnamefont {Da}}, \bibinfo
  {author} {\bibfnamefont {J.}~\bibnamefont {Joo}}, \bibinfo {author}
  {\bibfnamefont {E.}~\bibnamefont {Grinspun}},\ and\ \bibinfo {author}
  {\bibfnamefont {P.~M.}\ \bibnamefont {Reis}},\ }\bibfield  {title} {\bibinfo
  {title} {Coiling of elastic rods on rigid substrates},\ }\href@noop {}
  {\bibfield  {journal} {\bibinfo  {journal} {Proceedings of the National
  Academy of Sciences}\ ,\ \bibinfo {pages} {201409118}} (\bibinfo {year}
  {2014})}\BibitemShut {NoStop}%
\bibitem [{\citenamefont {Taylor}(1969)}]{taylor1969instability}%
  \BibitemOpen
  \bibfield  {author} {\bibinfo {author} {\bibfnamefont {G.~I.}\ \bibnamefont
  {Taylor}},\ }\bibfield  {title} {\bibinfo {title} {Instability of jets,
  threads, and sheets of viscous fluid},\ }in\ \href@noop {} {\emph {\bibinfo
  {booktitle} {Proc. Intl. Congress on Theor. Appl. Mechanics}}}\ (\bibinfo
  {publisher} {Springer},\ \bibinfo {year} {1969})\ pp.\ \bibinfo {pages}
  {382--388}\BibitemShut {NoStop}%
\bibitem [{\citenamefont {Mahadevan}\ \emph {et~al.}(1998)\citenamefont
  {Mahadevan}, \citenamefont {Ryu},\ and\ \citenamefont
  {Samuel}}]{mahadevan1998fluid}%
  \BibitemOpen
  \bibfield  {author} {\bibinfo {author} {\bibfnamefont {L.}~\bibnamefont
  {Mahadevan}}, \bibinfo {author} {\bibfnamefont {W.~S.}\ \bibnamefont {Ryu}},\
  and\ \bibinfo {author} {\bibfnamefont {A.~D.}\ \bibnamefont {Samuel}},\
  }\bibfield  {title} {\bibinfo {title} {Fluid ‘rope trick’investigated},\
  }\href@noop {} {\bibfield  {journal} {\bibinfo  {journal} {Nature}\ }\textbf
  {\bibinfo {volume} {392}},\ \bibinfo {pages} {140} (\bibinfo {year}
  {1998})}\BibitemShut {NoStop}%
\bibitem [{\citenamefont {Mahadevan}\ \emph {et~al.}(2000)\citenamefont
  {Mahadevan}, \citenamefont {Ryu},\ and\ \citenamefont
  {Samuel}}]{mahadevan2000correction}%
  \BibitemOpen
  \bibfield  {author} {\bibinfo {author} {\bibfnamefont {L.}~\bibnamefont
  {Mahadevan}}, \bibinfo {author} {\bibfnamefont {W.}~\bibnamefont {Ryu}},\
  and\ \bibinfo {author} {\bibfnamefont {A.}~\bibnamefont {Samuel}},\
  }\bibfield  {title} {\bibinfo {title} {Correction: fluid ‘rope
  trick’investigated},\ }\href@noop {} {\bibfield  {journal} {\bibinfo
  {journal} {Nature}\ }\textbf {\bibinfo {volume} {403}},\ \bibinfo {pages}
  {1038} (\bibinfo {year} {2000})}\BibitemShut {NoStop}%
\bibitem [{\citenamefont {Ribe}(2017)}]{ribe2017liquid}%
  \BibitemOpen
  \bibfield  {author} {\bibinfo {author} {\bibfnamefont {N.~M.}\ \bibnamefont
  {Ribe}},\ }\bibfield  {title} {\bibinfo {title} {Liquid rope coiling: a
  synoptic view},\ }\href@noop {} {\bibfield  {journal} {\bibinfo  {journal}
  {Journal of Fluid Mechanics}\ }\textbf {\bibinfo {volume} {812}} (\bibinfo
  {year} {2017})}\BibitemShut {NoStop}%
\bibitem [{\citenamefont {Kim}\ \emph {et~al.}(2010)\citenamefont {Kim},
  \citenamefont {Lee}, \citenamefont {Park}, \citenamefont {Kim},\ and\
  \citenamefont {Mahadevan}}]{kim2010nanopottery}%
  \BibitemOpen
  \bibfield  {author} {\bibinfo {author} {\bibfnamefont {H.-Y.}\ \bibnamefont
  {Kim}}, \bibinfo {author} {\bibfnamefont {M.}~\bibnamefont {Lee}}, \bibinfo
  {author} {\bibfnamefont {K.~J.}\ \bibnamefont {Park}}, \bibinfo {author}
  {\bibfnamefont {S.}~\bibnamefont {Kim}},\ and\ \bibinfo {author}
  {\bibfnamefont {L.}~\bibnamefont {Mahadevan}},\ }\bibfield  {title} {\bibinfo
  {title} {Nanopottery: coiling of electrospun polymer nanofibers},\
  }\href@noop {} {\bibfield  {journal} {\bibinfo  {journal} {Nano letters}\
  }\textbf {\bibinfo {volume} {10}},\ \bibinfo {pages} {2138} (\bibinfo {year}
  {2010})}\BibitemShut {NoStop}%
\bibitem [{\citenamefont {Brun}\ \emph {et~al.}(2017)\citenamefont {Brun},
  \citenamefont {Inamura}, \citenamefont {Lizardo}, \citenamefont {Franchin},
  \citenamefont {Stern}, \citenamefont {Houk},\ and\ \citenamefont
  {Oxman}}]{brun2017molten}%
  \BibitemOpen
  \bibfield  {author} {\bibinfo {author} {\bibfnamefont {P.-T.}\ \bibnamefont
  {Brun}}, \bibinfo {author} {\bibfnamefont {C.}~\bibnamefont {Inamura}},
  \bibinfo {author} {\bibfnamefont {D.}~\bibnamefont {Lizardo}}, \bibinfo
  {author} {\bibfnamefont {G.}~\bibnamefont {Franchin}}, \bibinfo {author}
  {\bibfnamefont {M.}~\bibnamefont {Stern}}, \bibinfo {author} {\bibfnamefont
  {P.}~\bibnamefont {Houk}},\ and\ \bibinfo {author} {\bibfnamefont
  {N.}~\bibnamefont {Oxman}},\ }\bibfield  {title} {\bibinfo {title} {The
  molten glass sewing machine},\ }\href@noop {} {\bibfield  {journal} {\bibinfo
   {journal} {Philosophical Transactions of the Royal Society A: Mathematical,
  Physical and Engineering Sciences}\ }\textbf {\bibinfo {volume} {375}},\
  \bibinfo {pages} {20160156} (\bibinfo {year} {2017})}\BibitemShut {NoStop}%
\bibitem [{\citenamefont {Dilworth}\ and\ \citenamefont
  {Bogue}(2015)}]{dilworth2015hand}%
  \BibitemOpen
  \bibfield  {author} {\bibinfo {author} {\bibfnamefont {P.}~\bibnamefont
  {Dilworth}}\ and\ \bibinfo {author} {\bibfnamefont {M.}~\bibnamefont
  {Bogue}},\ }\href@noop {} {\bibinfo {title} {Hand-held three-dimensional
  drawing device}} (\bibinfo {year} {2015}),\ \bibinfo {note} {uS Patent
  9,102,098}\BibitemShut {NoStop}%
\bibitem [{\citenamefont {Lin}\ and\ \citenamefont
  {Rutledge}(2018)}]{Rutledge}%
  \BibitemOpen
  \bibfield  {author} {\bibinfo {author} {\bibfnamefont {Y.-M.}\ \bibnamefont
  {Lin}}\ and\ \bibinfo {author} {\bibfnamefont {G.~C.}\ \bibnamefont
  {Rutledge}},\ }\bibfield  {title} {\bibinfo {title} {Separation of
  oil-in-water emulsions stabilized by different types of surfactants using
  electrospun fiber membranes},\ }\href@noop {} {\bibfield  {journal} {\bibinfo
   {journal} {Journal of membrane science}\ }\textbf {\bibinfo {volume}
  {563}},\ \bibinfo {pages} {247} (\bibinfo {year} {2018})}\BibitemShut
  {NoStop}%
\bibitem [{\citenamefont {Lee}\ and\ \citenamefont
  {Rogers}(2012)}]{lee2012effect}%
  \BibitemOpen
  \bibfield  {author} {\bibinfo {author} {\bibfnamefont {P.}~\bibnamefont
  {Lee}}\ and\ \bibinfo {author} {\bibfnamefont {M.}~\bibnamefont {Rogers}},\
  }\bibfield  {title} {\bibinfo {title} {Effect of calcium source and
  exposure-time on basic caviar spherification using sodium alginate},\
  }\href@noop {} {\bibfield  {journal} {\bibinfo  {journal} {International
  Journal of Gastronomy and Food Science}\ }\textbf {\bibinfo {volume} {1}},\
  \bibinfo {pages} {96} (\bibinfo {year} {2012})}\BibitemShut {NoStop}%
\bibitem [{\citenamefont {Machin}(1958)}]{machin1958wave}%
  \BibitemOpen
  \bibfield  {author} {\bibinfo {author} {\bibfnamefont {K.}~\bibnamefont
  {Machin}},\ }\bibfield  {title} {\bibinfo {title} {Wave propagation along
  flagella},\ }\href@noop {} {\bibfield  {journal} {\bibinfo  {journal}
  {Journal of Experimental Biology}\ }\textbf {\bibinfo {volume} {35}},\
  \bibinfo {pages} {796} (\bibinfo {year} {1958})}\BibitemShut {NoStop}%
\bibitem [{\citenamefont {Gosselin}\ \emph {et~al.}(2014)\citenamefont
  {Gosselin}, \citenamefont {Neetzow},\ and\ \citenamefont
  {Paak}}]{gosselin2014buckling}%
  \BibitemOpen
  \bibfield  {author} {\bibinfo {author} {\bibfnamefont {F.}~\bibnamefont
  {Gosselin}}, \bibinfo {author} {\bibfnamefont {P.}~\bibnamefont {Neetzow}},\
  and\ \bibinfo {author} {\bibfnamefont {M.}~\bibnamefont {Paak}},\ }\bibfield
  {title} {\bibinfo {title} {Buckling of a beam extruded into highly viscous
  fluid},\ }\href@noop {} {\bibfield  {journal} {\bibinfo  {journal} {Physical
  Review E}\ }\textbf {\bibinfo {volume} {90}},\ \bibinfo {pages} {052718}
  (\bibinfo {year} {2014})}\BibitemShut {NoStop}%
\bibitem [{\citenamefont {Cox}(1970)}]{cox_1970}%
  \BibitemOpen
  \bibfield  {author} {\bibinfo {author} {\bibfnamefont {R.~G.}\ \bibnamefont
  {Cox}},\ }\bibfield  {title} {\bibinfo {title} {The motion of long slender
  bodies in a viscous fluid part 1. general theory},\ }\href
  {https://doi.org/10.1017/S002211207000215X} {\bibfield  {journal} {\bibinfo
  {journal} {Journal of Fluid Mechanics}\ }\textbf {\bibinfo {volume} {44}},\
  \bibinfo {pages} {791–810} (\bibinfo {year} {1970})}\BibitemShut {NoStop}%
\bibitem [{\citenamefont {Andrews}\ \emph {et~al.}(1999)\citenamefont
  {Andrews}, \citenamefont {Askey},\ and\ \citenamefont
  {Roy}}]{special-functions}%
  \BibitemOpen
  \bibfield  {author} {\bibinfo {author} {\bibfnamefont {G.~E.}\ \bibnamefont
  {Andrews}}, \bibinfo {author} {\bibfnamefont {R.}~\bibnamefont {Askey}},\
  and\ \bibinfo {author} {\bibfnamefont {R.}~\bibnamefont {Roy}},\ }\href@noop
  {} {\emph {\bibinfo {title} {Special functions}}},\ Vol.~\bibinfo {volume}
  {71}\ (\bibinfo  {publisher} {Cambridge university press},\ \bibinfo {year}
  {1999})\BibitemShut {NoStop}%
\bibitem [{\citenamefont {Skylar-Scott}\ \emph {et~al.}(2019)\citenamefont
  {Skylar-Scott}, \citenamefont {Mueller}, \citenamefont {Visser},\ and\
  \citenamefont {Lewis}}]{skylar2019voxelated}%
  \BibitemOpen
  \bibfield  {author} {\bibinfo {author} {\bibfnamefont {M.~A.}\ \bibnamefont
  {Skylar-Scott}}, \bibinfo {author} {\bibfnamefont {J.}~\bibnamefont
  {Mueller}}, \bibinfo {author} {\bibfnamefont {C.~W.}\ \bibnamefont
  {Visser}},\ and\ \bibinfo {author} {\bibfnamefont {J.~A.}\ \bibnamefont
  {Lewis}},\ }\bibfield  {title} {\bibinfo {title} {Voxelated soft matter via
  multimaterial multinozzle 3d printing},\ }\href@noop {} {\bibfield  {journal}
  {\bibinfo  {journal} {Nature}\ }\textbf {\bibinfo {volume} {575}},\ \bibinfo
  {pages} {330} (\bibinfo {year} {2019})}\BibitemShut {NoStop}%
\bibitem [{\citenamefont {Style}\ \emph {et~al.}(2020)\citenamefont {Style},
  \citenamefont {Tutika}, \citenamefont {Kim},\ and\ \citenamefont
  {Bartlett}}]{style2020solid}%
  \BibitemOpen
  \bibfield  {author} {\bibinfo {author} {\bibfnamefont {R.~W.}\ \bibnamefont
  {Style}}, \bibinfo {author} {\bibfnamefont {R.}~\bibnamefont {Tutika}},
  \bibinfo {author} {\bibfnamefont {J.~Y.}\ \bibnamefont {Kim}},\ and\ \bibinfo
  {author} {\bibfnamefont {M.~D.}\ \bibnamefont {Bartlett}},\ }\bibfield
  {title} {\bibinfo {title} {Solid--liquid composites for soft multifunctional
  materials},\ }\href@noop {} {\bibfield  {journal} {\bibinfo  {journal}
  {Advanced Functional Materials}\ ,\ \bibinfo {pages} {2005804}} (\bibinfo
  {year} {2020})}\BibitemShut {NoStop}%
\bibitem [{\citenamefont {Bartlett}\ and\ \citenamefont
  {Style}(2020)}]{bartlett2020introduction}%
  \BibitemOpen
  \bibfield  {author} {\bibinfo {author} {\bibfnamefont {M.~D.}\ \bibnamefont
  {Bartlett}}\ and\ \bibinfo {author} {\bibfnamefont {R.~W.}\ \bibnamefont
  {Style}},\ }\bibfield  {title} {\bibinfo {title} {Introduction to liquid
  composites},\ }\href@noop {} {\bibfield  {journal} {\bibinfo  {journal} {Soft
  Matter}\ }\textbf {\bibinfo {volume} {16}},\ \bibinfo {pages} {5799}
  (\bibinfo {year} {2020})}\BibitemShut {NoStop}%
\bibitem [{\citenamefont {Cai}\ \emph {et~al.}(2019)\citenamefont {Cai},
  \citenamefont {Marthelot},\ and\ \citenamefont {Brun}}]{cai2019unbounded}%
  \BibitemOpen
  \bibfield  {author} {\bibinfo {author} {\bibfnamefont {L.}~\bibnamefont
  {Cai}}, \bibinfo {author} {\bibfnamefont {J.}~\bibnamefont {Marthelot}},\
  and\ \bibinfo {author} {\bibfnamefont {P.-T.}\ \bibnamefont {Brun}},\
  }\bibfield  {title} {\bibinfo {title} {An unbounded approach to microfluidics
  using the {R}ayleigh {P}lateau instability of viscous threads directly drawn
  in a bath},\ }\href@noop {} {\bibfield  {journal} {\bibinfo  {journal}
  {Proceedings of the National Academy of Sciences}\ }\textbf {\bibinfo
  {volume} {116}},\ \bibinfo {pages} {22966} (\bibinfo {year}
  {2019})}\BibitemShut {NoStop}%
\bibitem [{\citenamefont {Onoe}\ \emph {et~al.}(2013)\citenamefont {Onoe},
  \citenamefont {Okitsu}, \citenamefont {Itou}, \citenamefont {Kato-Negishi},
  \citenamefont {Gojo}, \citenamefont {Kiriya}, \citenamefont {Sato},
  \citenamefont {Miura}, \citenamefont {Iwanaga}, \citenamefont
  {Kuribayashi-Shigetomi} \emph {et~al.}}]{onoe2013metre}%
  \BibitemOpen
  \bibfield  {author} {\bibinfo {author} {\bibfnamefont {H.}~\bibnamefont
  {Onoe}}, \bibinfo {author} {\bibfnamefont {T.}~\bibnamefont {Okitsu}},
  \bibinfo {author} {\bibfnamefont {A.}~\bibnamefont {Itou}}, \bibinfo {author}
  {\bibfnamefont {M.}~\bibnamefont {Kato-Negishi}}, \bibinfo {author}
  {\bibfnamefont {R.}~\bibnamefont {Gojo}}, \bibinfo {author} {\bibfnamefont
  {D.}~\bibnamefont {Kiriya}}, \bibinfo {author} {\bibfnamefont
  {K.}~\bibnamefont {Sato}}, \bibinfo {author} {\bibfnamefont {S.}~\bibnamefont
  {Miura}}, \bibinfo {author} {\bibfnamefont {S.}~\bibnamefont {Iwanaga}},
  \bibinfo {author} {\bibfnamefont {K.}~\bibnamefont {Kuribayashi-Shigetomi}},
  \emph {et~al.},\ }\bibfield  {title} {\bibinfo {title} {Metre-long cell-laden
  microfibres exhibit tissue morphologies and functions},\ }\href@noop {}
  {\bibfield  {journal} {\bibinfo  {journal} {Nature materials}\ }\textbf
  {\bibinfo {volume} {12}},\ \bibinfo {pages} {584} (\bibinfo {year}
  {2013})}\BibitemShut {NoStop}%
\bibitem [{\citenamefont {Hu}\ \emph {et~al.}(2010)\citenamefont {Hu},
  \citenamefont {Pasta}, \citenamefont {La~Mantia}, \citenamefont {Cui},
  \citenamefont {Jeong}, \citenamefont {Deshazer}, \citenamefont {Choi},
  \citenamefont {Han},\ and\ \citenamefont {Cui}}]{hu2010stretchable}%
  \BibitemOpen
  \bibfield  {author} {\bibinfo {author} {\bibfnamefont {L.}~\bibnamefont
  {Hu}}, \bibinfo {author} {\bibfnamefont {M.}~\bibnamefont {Pasta}}, \bibinfo
  {author} {\bibfnamefont {F.}~\bibnamefont {La~Mantia}}, \bibinfo {author}
  {\bibfnamefont {L.}~\bibnamefont {Cui}}, \bibinfo {author} {\bibfnamefont
  {S.}~\bibnamefont {Jeong}}, \bibinfo {author} {\bibfnamefont {H.~D.}\
  \bibnamefont {Deshazer}}, \bibinfo {author} {\bibfnamefont {J.~W.}\
  \bibnamefont {Choi}}, \bibinfo {author} {\bibfnamefont {S.~M.}\ \bibnamefont
  {Han}},\ and\ \bibinfo {author} {\bibfnamefont {Y.}~\bibnamefont {Cui}},\
  }\bibfield  {title} {\bibinfo {title} {Stretchable, porous, and conductive
  energy textiles},\ }\href@noop {} {\bibfield  {journal} {\bibinfo  {journal}
  {Nano letters}\ }\textbf {\bibinfo {volume} {10}},\ \bibinfo {pages} {708}
  (\bibinfo {year} {2010})}\BibitemShut {NoStop}%
\end{thebibliography}%

\end{document}